\documentclass[twocolumn,floatfix,aps,showpacs,showkeys]{revtex4-1}
\usepackage{epsfig,subfigure,latexsym}

\usepackage{amssymb,amsmath}

\begin{document}
\def\Journal#1#2#3#4{{\it #1} {\bf #2}, #3 (#4) }

\title{Quark Matter at High Density based on Extended Confined-isospin-density-dependent-mass Model}

\author{A. I. Qauli, and A. Sulaksono}

\affiliation{Departemen Fisika, FMIPA, Universitas Indonesia, Depok 16424, Indonesia }

\begin{abstract}  
We investigate the effect of the inclusion of relativistic Coulomb terms in a confined-isospin-density-dependent-mass (CIDDM) model of strange quark matter (SQM). We found that if we include Coulomb term in scalar density form, SQM equation of state (EOS)  at high densities is stiffer but if we include Coulomb term in vector  density form is softer than that of  standard CIDDM model. We also investigate systematically the role of each term of the extended CIDDM model. Compared with what was reported in Ref.~\cite {ref:isospin}, we found the stiffness of SQM EOS is controlled by the interplay among the the oscillator harmonic, isospin asymmetry and Coulomb contributions depending on the parameter's range of these terms. We have found that the absolute stable condition of SQM and the mass of 2 $M_\odot$ pulsars can constrain the parameter of  oscillator harmonic $\kappa_1$ $\approx 0.53$ in the case Coulomb term excluded. If the Coulomb term is included, for the models with their parameters are consistent with SQM absolute stability condition, the  $2.0~M_{\odot}$ constraint more prefer the maximum mass prediction of model with scalar Coulomb term than that of model with vector Coulomb term. On contrary, the high densities EOS predicted by  model with vector Coulomb is more compatible with recent pQCD result \cite{ref:pressure} than that predicted by model with scalar Coulomb. Furthermore, we also observed the quark composition in a very high density region depends quite sensitively on the kind of Coulomb term used.
\end{abstract} 

\keywords{Strange quark star, Pulsar, equation of state} 
\pacs{12.39.-x,12.39.Ki,21.65.Qr,97.60.Gb}

\maketitle
 \section{INTRODUCTION}
\label{sec_intro}
Recently, the mass 1.97 $\pm$ 0.04 $M_\odot$ of pulsar J1614-2230 is measured from the Shapiro delay~\cite {Demorest10} and the mass 2.01 $\pm$ 0.04 $M_\odot$~\cite{Antoniadis13} of pulsar J0348+0432 is also measured from the gravitational redshift optical lines of its  white dwarf companion. These measurements put restricted constraints on compact star EOS. Pulsar has long been proposed to be neutron star (see review for examples Refs~\cite{MCM2013,CHZF2013,LP2010}) or hybrid star where the quark matter can exist in the core of neutron star (See Ref.\cite{ref:hybrid} and references therein). Another possibility is that pulsar is strange stars i.e., stellar object, composed entirely of strange quark matter (SQM). The possibility and the implications of the idea that pulsars are born as strange stars are explored by the authors of Ref.~\cite{ref:strange} (see also the references therein). In addition, it is reported that PSR 0943+10 fits neither in the category of neutron star nor in that of black holes and the apparent compactness of this pulsar could be explained only if it is composed of quarks~\cite{ref:strange2}.  Furthermore, it is also important to point out that the authors  of Ref.~\cite{AAS2014} demonstrated through a detailed analysis of the pulsar evolution how precise pulsar timing data can constrain the star's composition. They also found that interacting quark matter is consistent with both the observed radio and x-ray data, whereas for ordinary nuclear matter, additional enhanced damping mechanism will be required. However, up to now, we cannot determine surely whether the pulsar is neutron star, hybrid star,  or strange quark star only from looking at the mass-radius relation\cite{ref:hybrid}. Not only neutron stars, but also pure quark or hybrid stars can yield a mass  $\ge 2.0M_\odot$. Moreover, it is reported~\cite{ref:cold} that purely quark matter stars can yield mass $\ge 2.5M_\odot$. We need to point out here that SQM is more stable than nuclear matter \cite{ref:strange}. The stability of SQM is determined from minimum energy per baryon. When it is less than the mass of $^{56}$Fe, then SQM is absolutely stable. When it is larger than the mass of  $^{56}$Fe but still less than nucleon mass then it becomes metastable. Also, at the same time the minimum energy per baryon of beta-equilibrium two-flavor \textit{u-d} quark matter should be larger than $930~\rm{MeV}$. Otherwise, SQM becomes unstable~\cite{ref:Farhi}. It is also important to note that at high density and low temperature, hybrid stars generally predict smaller mass than the neutron star composed entirely of hadronic matter~\cite{ref:weissenborn}.
 
The MIT Bag model is a phenomenological approach that provides the simplest description for stable SQM. This model later has been modified by adding density dependence to the quark's mass term instead of using a constant bag to simulate the interactions. This model is widely known as the confined-density-dependent-mass (CDDM) model (see Ref.\cite{ref:cddm} and references therein). Recently, an extended version of the CDDM model i.e., by including an isospin interaction that is named by the authors as the confined-isospin-density-dependent-mass (CIDDM) model, is proposed~\cite{ref:isospin}. The authors~\cite{ref:isospin} have reported the CIDDM model with strong isospin asymmetry dependence can yield  strange star mass of $2.0M_\odot$. This result strengthens the possibility that the existence of strange stars still cannot be ruled out. We note the effect of a strong magnetic field in a strange star using the CIDDM model as the framework has been investigated by the authors of Ref.~\cite{Cu2014}, and besides MIT Bag and CIDDM, there are other phenomenological models, such as the Nambu-Jona-Lasinio (NJL) model~\cite{NJL}, Dyson-Schwinger approach~\cite{DSA}, and pQCD approach~\cite{ref:cold,ref:pressure,ref:pressure2}, have been used to study SQM.

In this work, we extend the CIDDM model by adding relativistic Coulomb terms. One of these Coulomb terms (the scalar density form) also have been investigated recently by the authors of Ref.\cite{ref:coulomb}. However, they focus more on studying the thermodynamics consistency description of SQM. Here, our aim is to find out which kind of Coulomb term is more appropriate for SQM by studying their effects on absolute stability condition of SQM, also on EOS at high densities and mass-radius relation of the star. We found the value of the corresponding  parameter of Coulomb term in a form of a vector density can be adjusted, thus the EOS of SQM at high density can be compatible with the one predicted recently by using the pQCD approach\cite{ref:pressure}. However, we need to pay attention at high densities or large chemical potential ($\mu_B$ $>$ 4 GeV) where the charm quarks must appear but neglected in Ref.\cite{ref:pressure}. On the other hand, we have found also that Coulomb term in a form of scalar density is more compatible with  $2.0M_\odot$ constraint. We also obtained that the stiffness of SQM EOS is controlling by the interplay among the the oscillator harmonic, isospin asymmetry and Coulomb terms depending on the parameters range of these terms.  Moreover, we show that at high densities, the role of Coulomb terms becomes more crucial  in determining the EOS of SQM than that at low and moderate  densities.  In our calculation, the best parameter set from Ref.\cite{ref:isospin}, which yields a quark star mass of $2.0M_\odot$, is used as the basis and then we explore the SQM properties by using several variations of parameters  of the extended CIDDM model.

\section{FORMALISM}
\label{sec_Formalism}
In this section, we briefly review the formalism of the CIDDM model with additional Coulomb terms. In the original CIDDM model presented in Ref.~\cite {ref:isospin}, the quark interactions are modeled by assuming the quark masses are density- and isospin-dependent. This formalism can also be expressed equivalently by writing the Lagrangian density of CIDDM as
\begin{eqnarray}
\mathcal{L}^{\rm{CIDDM}}=\mathcal{L}^{\rm{K}}+\mathcal{L}^{\rm{Int({CIDDM})}},
\end{eqnarray}
where Lagrangian for the free quarks kinetic term is
\begin{eqnarray}
\mathcal{L}^{\rm{K}}=\sum_{j=u,d,s}{\frac{g_j}{\left({2 \pi}\right)^3} \int_{0}^{{k_F}_j} \bar{\psi_j}\left({i \gamma^{\mu} \partial_{\mu} - m_{j0} }\right)\psi_j d^3 k}, \nonumber \\
\end{eqnarray}
where  $m_{u0}$, $m_{d0}$ are mass of up and down quarks, respectively,  and $m_{s0}$ is the mass of strange quark. For the interaction terms, which simulate the quark confinement, can be presented as
\begin{eqnarray}
\mathcal{L}^{\rm{Int({CIDDM})}}&\equiv&-\kappa_1 {n_B}^{-1/3} {n_B}^{(s)}-\kappa_{3}\delta {n_{B}}^{a}e^{-b n_{B}}{n_B}^{(s\tau)}. \nonumber \\
\label{INTLAG1}
\end{eqnarray} 

Here, $\kappa_1$ and $\kappa_3$ are quark's isospin independent and dependent interaction parameters, while $a$ and $b$ in the second term are isospin parameters in the CIDDM model. $\delta$ in Eq. (\ref{INTLAG1}) denotes the up-down quark asymmetry parameter, which is defined as
\begin{eqnarray}
\delta=3\left({n_d-n_u}\right)/\left({n_d+n_u}\right), 
\end{eqnarray} 
where  the number density of $j$ quark $n_j$ is ${{k_F}_j}^3/\pi^2$. Vector-isoscalar, scalar-isoscalar, and vector-isovector baryon densities  $n_B$, ${n_B}^{s}$, and ${n_B}^{s\tau}$  in Eq. (\ref{INTLAG1}) are defined as
\begin{eqnarray}
n_B&=& \frac{n_u + n_d + n_s}{3}, \\
{n_B}^{(s)}&=&\sum_{j=u,d,s}{\frac{g_j}{\left({2 \pi}\right)^3} \int_{0}^{{k_F}_j} \bar{\psi_j}}\psi_{j}d^3k, \\
{n_B}^{(s\tau)}&=&\sum_{j=u,d,s}{\frac{g_j}{\left({2 \pi}\right)^3} \int_{0}^{{k_F}_j}\tau_j \bar{\psi_j}}\psi_{j}d^3k.
\end{eqnarray}
 $\tau_j$ is the isospin quantum number of quarks, where $\tau_j = 1$ for $j=u$ (up quarks), $\tau_j = -1$ for $j=d$ (down quarks), and $\tau_j=0$ for $j=s$ (strange quarks).

In this work, we focus on investigating the Lagrangian density interaction in the following form 
\begin{eqnarray}
\mathcal{L}^{\rm{Int}}=\mathcal{L}^{\rm{Int(CIDDM)}}+\Delta \mathcal{L}^{(i)\rm{Int}},
\label{LagInt}
\end{eqnarray}
 where here, we call the model with Coulomb scalar density i.e.,
\begin{eqnarray}
\Delta\mathcal{L}^{(1)\rm{Int}}\equiv-{\kappa_2}^{(1)}{n_B}^{1/3}{n_B}^{(s)}
\end{eqnarray}
 as model I and the model with Coulomb vector density i.e., 
\begin{eqnarray}
\Delta \mathcal{L}^{(2)\rm{Int}}\equiv-{\kappa_2}^{(2)}{n_B}^{4/3}
\end{eqnarray}
 as model II. It can be seen that $\Delta \mathcal{L}^{(1)\rm{Int}}$ and $\Delta \mathcal{L}^{(2)\rm{Int}}$ behave differently, only in high density limit. However, in low density limit, i.e., when $k_F\to 0$, then $\Delta \mathcal{L}^{(1)\rm{Int}}\approx \Delta \mathcal{L}^{(2)\rm{Int}}$. We can see the physical meaning of each term in  Eq.~(\ref{LagInt}) easily if we neglect for a while the isospin-dependent term ($\kappa_3$ term) in Lagrangian density in Eq. ~(\ref{LagInt}) and taking non-relativistic limit, i.e., $k_F\to0$ of $\mathcal{L}^{\rm{Int}}$. In this limit, the interaction potential per baryon can be obtained from the following expression 
\begin{eqnarray}
V(r)\approx-{n_B}^{-1}\mathcal{L}^{\rm{Int}},
\end{eqnarray} 
and further, if we use $r\propto {n_B}^{-1/3}$, then it is clear that $V(r)$ becomes
\begin{eqnarray}
V(r)\propto \kappa_1 r^2 + {\kappa_2}^{(1 ~or~ 2)}r^{-1},
\end{eqnarray}
which is known as Cornell Potential~\cite{ref:Cornell}. Therefore, we can interpret the term behind parameter $\kappa_1$ as a harmonic oscillator term and the term behind ${\kappa_2}^{(1 ~or~ 2)}$ as a Coulomb term. It is interesting to compare the energy density behavior at high densities of model I and model II. Model I has energy density as
\begin{eqnarray}
\epsilon^{(1)}=\sum_j \frac{g_j}{\left({2 \pi}\right)^3} \int_{0}^{{k_F}_j} \bar{\psi_j}\left[{\gamma^0 \left({\hat{\alpha}.\hat{p}}\right)+{m_j}^{\left({1}\right)}}\right]\psi_j d^3k, \label{eq:energy1} \nonumber \\
\end{eqnarray}
with density- and isospin-dependent quark masses as
\begin{eqnarray}
{m_j}^{\left({1}\right)}&=&m_{j0}+\kappa_1 {n_B}^{-1/3} +\kappa_2^{(1)} {n_B}^{1/3}  \nonumber \\
&&+\tau_j \kappa_{3} \delta{n_{B}}^{a}e^{-b n_{B}}, \label{eq:mass1}
\end{eqnarray}

while model II on the other hand yields
\begin{eqnarray}
\epsilon^{(2)}&=&\sum_j \frac{g_j}{\left({2 \pi}\right)^3} \int_{0}^{{k_F}_j} \bar{\psi_j}\left[{\gamma^0 \left({\hat{\alpha}.\hat{p}}\right)+{m_j}^{(2)}}\right]\psi_j d^3k \nonumber \\
&&+\kappa_2^{(2)} {n_B}^{4/3}, \label{eq:energy2}
\end{eqnarray}
with density- and isospin-dependent quark masses equivalent to the original CIDDM model as
\begin{eqnarray}
{m_j}^{(2)}&=&m_{j0}+\kappa_1 {n_B}^{-1/3}+ \tau_j \kappa_{3} \delta{n_{B}}^{a}e^{-b n_{B}}.\label{eq:mass2}
\end{eqnarray}
Thus, it is obvious that the differences are due to the position of the Coulomb term in energy density. In model I, this term presents inside the terms of quark mass, while in model II, it appears outside the terms of quark mass. 

\section{RESULTS AND DISCUSSION}
\label{sec_randd}
\begin{figure}
\epsfig{figure=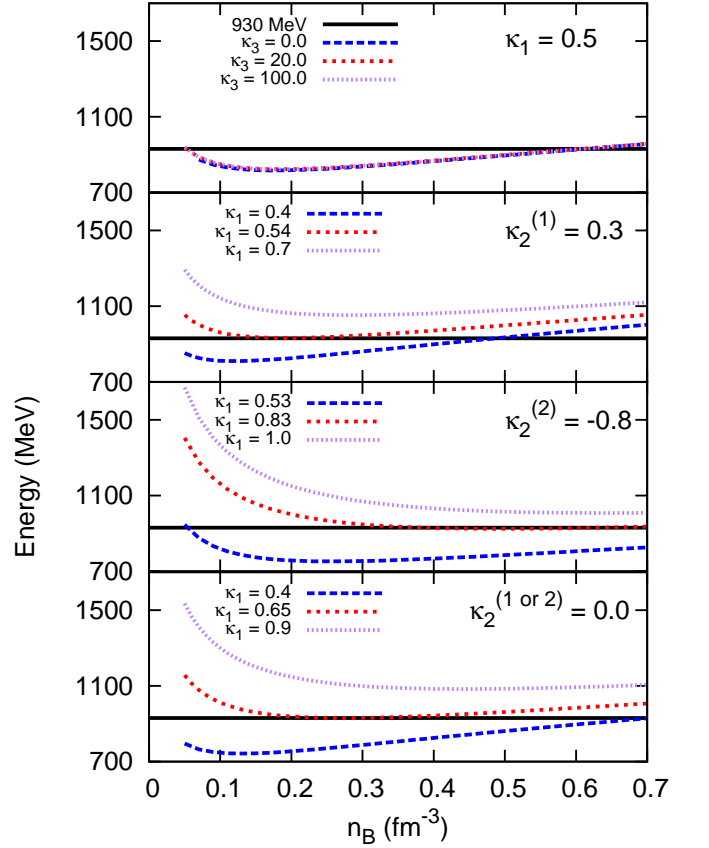, width=9cm}
\caption{\label{fig:energy} Energy per baryon for SQM as a function of baryon number density. For the lower panel (${\kappa_2}^{(1~\rm{or}~2)}=0$), the fixed parameters  are $\kappa_{2}^{(1~\rm{or}~2)}=0$, $\kappa_3=12.7$, $a=0.8$, and $b=0.1~\rm{fm}^{3}$. For the second lower panel (${\kappa_2}^{(2)}=-0.5$), the fixed parameters are ${\kappa_2}^{(1)}=0$, ${\kappa_2}^{(2)}=0.5$, $\kappa_3=12.7$, $a=0.8$, and $b=0.1~\rm{fm}^{3}$. For the second upper panel (${\kappa_2}^{(1)}=0.3$), the fixed parameters are $\kappa_{2}^{(1)}=0.3$, $\kappa_{2}^{(2)}=0$, $\kappa_{3}=12.7$, $a=0.8$, and $b=0.1~\rm{fm}^{3}$. For the upper panel ($\kappa_3$ variation), the fixed parameters  are $\kappa_1=0.5$, $\kappa_{2}^{(1~\rm{or}~2)}=0$, $a=0.8$, and $b=0.1 ~\rm{fm}^{3}$.}
\end{figure}  
\begin{figure}
\epsfig{figure=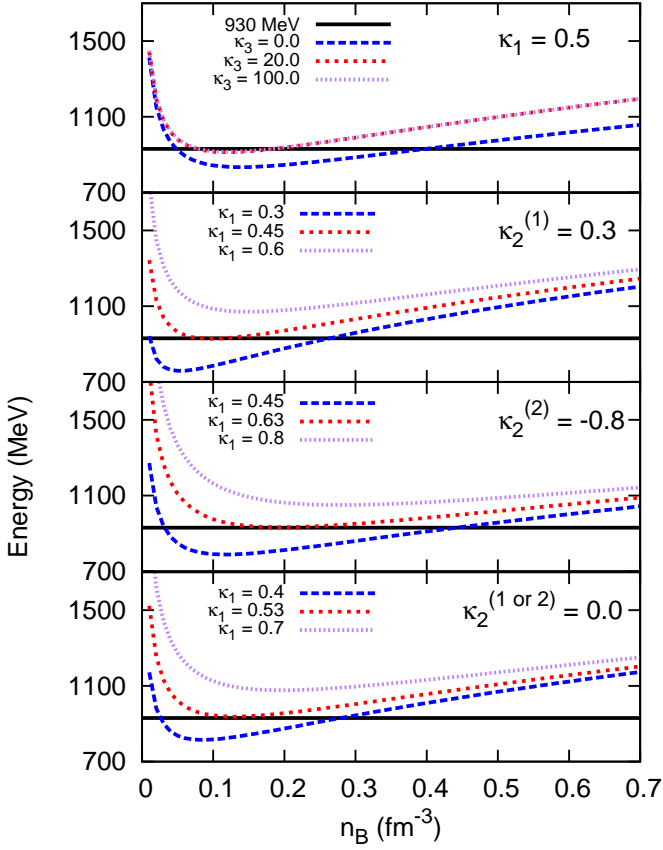, width=9cm}
\caption{\label{fig:energy_2} Energy per baryon for u-d quark matter as a function of baryon number density. The fixed parameter set used in each panel are the same as those used in Fig.~\ref{fig:energy}.}
\end{figure} 
In this section, we will discuss the consequences of the difference between model I and model II in EOS, SQM absolute stability condition and the mass-radius relation of a strange star.  We also discuss the role of each term in the extended CIDDM model. If we define pressure for quark matter as 
\begin{eqnarray}
P=-\epsilon+\sum\limits_{j=u,d,s,e}\mu_i n_i,
\end{eqnarray}
and for massless non-interacting quark matter\cite{ref:pressure,ref:cold} at high densities, it is known that
\begin{eqnarray}
P_{SB}=\frac{3}{4\pi^2}\left({\frac{\mu_B}{3}}\right)^4,
\end{eqnarray}
 where  the baryon chemical potential is defined as $\mu_B = \mu_u+\mu_d+\mu_s$.
Then, at high density, model I yields the pressure ratio:
\begin{eqnarray} 
\frac{P}{P_{SB}} &\approx& \frac{-C_1 + 3 C_2}{\frac{3}{4}\pi^{2/3}}, \label{eq:prediction1}
\end{eqnarray}
where 
\begin{eqnarray}
C_1&=&\frac{9}{8\pi^2} \left[{\pi^{2/3} \sqrt{\pi^{4/3} + {{\kappa_2}^{(1)}}^2} \left({2\pi^{4/3} + {{\kappa_2}^{(1)}}^2}\right) }\right. \nonumber \\
&&\left.{- {{\kappa_2}^{(1)}}^4 \ln{\frac{\pi^{2/3} + \sqrt{\pi^{4/3} + {{\kappa_2}^{(1)}}^2}}{{\kappa_2}^{(1)}} } }\right],
\end{eqnarray}
and
\begin{eqnarray}
C_2&=&\sqrt{\pi^{4/3} + {{\kappa_2}^{(1)}}^2}\nonumber \\
&&+\frac{{\kappa_2}^{(1)} }{2\pi^2} \left[{{\kappa_2}^{(1)} \pi^{2/3} \sqrt{\pi^{4/3} + {{\kappa_2}^{(1)}}^2}}\right. \nonumber \\
&&\left.{-{{\kappa_2}^{(1)}}^3\ln{\frac{\pi^{2/3} + \sqrt{\pi^{4/3} + {{\kappa_2}^{(1)}}^2}}{{\kappa_2}^{(1)}} }}\right].
\end{eqnarray}
It is clear if we set ${\kappa_2}^{(1)}$=0 that $\frac{P}{P_{SB}}\approx\left({\frac{P}{P_{SB}}}\right)_{0}$, where $\left({\frac{P}{P_{SB}}}\right)_{0}$ is the  pressure  at high density predicted by the original CIDDM model. This ratio can be approximated as
\begin{eqnarray}
\left({\frac{P}{P_{SB}}}\right)_0\approx 1.
\label{eq:predictionX}
\end{eqnarray} 
While on the other hand, model II yields
\begin{eqnarray}
\frac{P}{P_{SB}}\approx\left({\frac{P}{P_{SB}}}\right)_{0}+\frac{44}{3}\frac{{\kappa_2}^{(2)}}{\pi^{2/3}}, \label{eq:prediction2}
\end{eqnarray}
then, it is obvious if  we set ${\kappa_2}^{(2)}$=0 that $\frac{P}{P_{SB}}\approx\left({\frac{P}{P_{SB}}}\right)_{0}$. Eqs.~(\ref{eq:prediction1}) and~(\ref{eq:prediction2}) demonstrate that the role of Coulomb terms significantly appears only at high densities and the behavior of the EOS at high densities depends on whether we used vector (model II) or scalar (model I) densities to represent the  Coulomb interaction.

As we can see in Eq.~(\ref{eq:predictionX}), the $\frac{P}{P_{SB}}$ prediction from the CDDM or CIDDM is approximated as $1$ at high densities in which the Fermi momentum for each quark will be much larger than the quark's mass, whatever the value of the parameters used. On the other hand, the pQCD result without considering the presence of charm quarks shows the pressure of strange matter at high densities will be less than $1$\cite{ref:pressure,ref:cold}. This indicates that the actual $\frac{P}{P_{SB}}$ could be $\ne 1$. However, we have a greater degree of freedom by adding the Coulomb term to the CIDDM model so the  value of $\frac{P}{P_{SB}}$ at high densities can be adjusted. It is obvious by comparing the expressions in Eq.~(\ref{eq:prediction1}) and Eq.~(\ref{eq:prediction2}) that the vector Coulomb interaction in Eq.~(\ref{eq:prediction2}) provides more flexible form as the high densities correction term than that of Eq.~(\ref{eq:prediction1}) and we can easily adjust ${\kappa_2}^{(2)}$ to be compatible to the pQCD result\cite{ref:pressure,ref:cold}.  Furthermore, for the scalar Coulomb interaction in Eq.~(\ref{eq:prediction1}), if we assign $\frac{P}{P_{SB}}$ less than $1$, we found the solution for ${\kappa_2}^{(1)}$ is an unnatural complex number, while if we use a positive real number for ${\kappa_2}^{(1)}$, then $\frac{P}{P_{SB}} \ge 1$ at high densities. Note for the next, in calculating the EOS of strange star, we imposed charge neutrality and beta stability requirements and we use the unit of $\kappa_1$ in $\rm {fm}^{-2}$, $\kappa_3$ in $\rm {fm}^{3a-1}$, while ${\kappa_2}^{(1)}$ and ${\kappa_2}^{(2)}$ are dimensionless. Also, we assume $m_{\rm{u0}}=m_{\rm{d0}}=5.5~\rm{MeV}$ and $m_{\rm{s0}}=80~\rm{MeV}$.  

\begin{figure}
\epsfig{figure=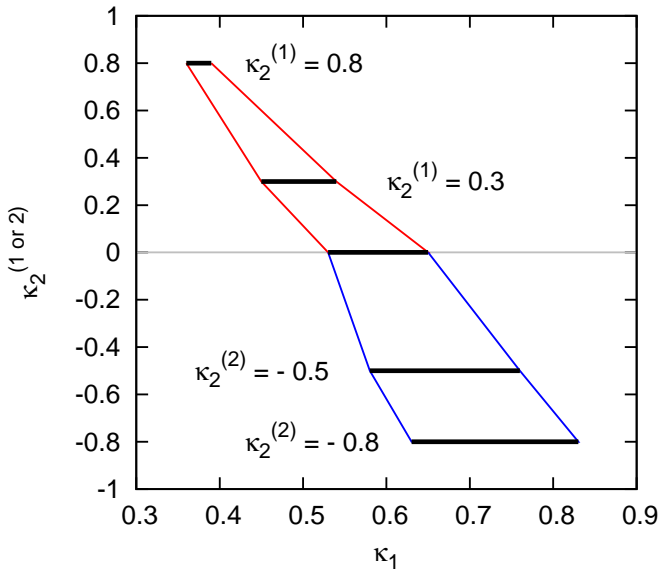, width=9cm}
\caption{\label{fig:beconstraint} $\kappa_{2}^{(1~\rm{or}~2)}$ as a function of allowed  $\kappa_1$ by absolute stability condition of SQM. The constraint of $\kappa_1$ (black line) are: $0.36 \lesssim \kappa_1 \lesssim 0.39$ for $\kappa_{2}^{(1)}=0.8$, $0.45 \lesssim \kappa_1 \lesssim 0.54$ for $\kappa_{2}^{(1)}=0.3$, $0.53 \lesssim \kappa_1 \lesssim 0.65$ for $\kappa_{2}^{(1~\rm{or}~2)}=0$, $0.58 \lesssim \kappa_1 \lesssim 0.76$ for $\kappa_2^{(2)}=-0.5$ and $0.63 \lesssim \kappa_1 \lesssim 0.83$ for $\kappa_{2}^{(2)}=-0.8$.}  
\end{figure}

In low density region, one of the most important point we need to investigate is the absolute stability condition of SQM.  The absolute stability condition can be reached when the minimum energy per baryon of SQM less than 930 MeV and at the same time the minimum energy per baryon of beta-equilibrium two-flavor \textit{u-d} quark matter should be larger than $930~\rm{MeV}$ \cite{ref:Farhi}. This condition can constrain the parameters of the extended CIDDM model. The results are shown in Fig.~\ref{fig:energy} and Fig.~\ref{fig:energy_2}. It can be seen from the lower panel of these figures that the value for $\kappa_1$ when Coulomb interaction excluded (${\kappa_2}^{(1~\rm{or}~2)}=0$) must be in the range of $0.53\lesssim \kappa_1 \lesssim 0.65$ in order the binding energies match with absolute stability condition. Also we can see from second lower panels of both figures that we can obtain the allowed range of $\kappa_1$ about $0.63 \lesssim \kappa_1 \lesssim 0.83$ if we used a fixed value of Coulomb parameter ${\kappa_2}^{(2)}=-0.8$. While from second upper panel of the figures, we can also obtain the allowed range of $\kappa_1$ about $0.45 \lesssim \kappa_1 \lesssim 0.54$ for a fixed value of  Coulomb parameter ${\kappa_2}^{(1)}=0.3$. It means that for the case of model I (with scalar Coulomb), by increasing the ${\kappa_2}^{(1)}$ causes the allowed range of  $\kappa_1$ narrowed and the range are shifted to the region with smaller value of  $\kappa_1$. Otherwise for the case of model II (with vector Coulomb), by increasing the ${\kappa_2}^{(1)}$ causes the allowed range of  $\kappa_1$ wider and the range are shifted to the region with larger value of  $\kappa_1$. As the consequence, the scalar Coulomb parameter tends to stiffen while vector Coulomb parameter tends to soften the EOS at high densities. This fact is explicitly shown in Fig.~\ref{fig:beconstraint}.    

From upper panel of Fig.~\ref{fig:energy} and Fig.~\ref{fig:energy_2}, we can also see clearly the variation of isospin parameter $\kappa_3$ provides insignificant effect in the EOS of SQM. However, this parameter yields pronounced effect in increasing the minimum energy per baryon in \textit{u-d} quark matter into the one which is consistent with the absolute stable condition constraint. On the other hand, the variation of isospin parameter $\kappa_3$ from $\kappa_{3}=20-100$ yields similar energy per baryon result for \textit{u-d} quark matter. Note, we show this behavior for the case $\kappa_ 1=0.5$, $a=0.8$, $b=0.1 ~\rm{fm}^{3}$ and $\kappa_{2}^{(1~\rm{or}~2)}=0$. Therefore, to constrain the upper bound of the isospin asymmetry parameter from absolute stability condition for SQM seems to be difficult. On the other hand, for the u-d quark matter on the upper panel of Fig~\ref{fig:energy_2}, the difference in minimum energy is only caused by "with"($\kappa_{3}=20-100$) and "without"($\kappa_{3}=0$) isospin asymmetry parameter. This fact provides lower bound constraint of $\kappa_{3} \gtrsim 20$ in order to fulfill the absolute stability condition constraint of binding energy of \textit{u-d} quark matter.

In general, the behavior of the EOS of SQM under CIDDM model depends on the interplay of three parameters (harmonic oscillator, isospin asymmetry, and Coulomb). However, from previous discussion, it seems the Coulomb parameter should be constrained by the information from high density properties, instead of the ones from low density. Therefore, if more realistic microscopic calculation of high density EOS of SQM, such as the one from pQCD but with including charm quarks at high densities, presents in the future, then we might decide which kind of Coulomb term is more appropriate as well as we might determine the exact value of the corresponding  Coulomb parameter. However, it is quite informative if we observe in this occasion the qualitative behavior of each parameter of extended  CIDDM model, by looking at the change of EOS at high density when the corresponding parameter of the model is varied, while other parameters are fixed, and compared to the pQCD EOS at high densities~\cite{ref:cold,ref:pressure,ref:pressure2}.  
\begin{figure}
\epsfig{figure=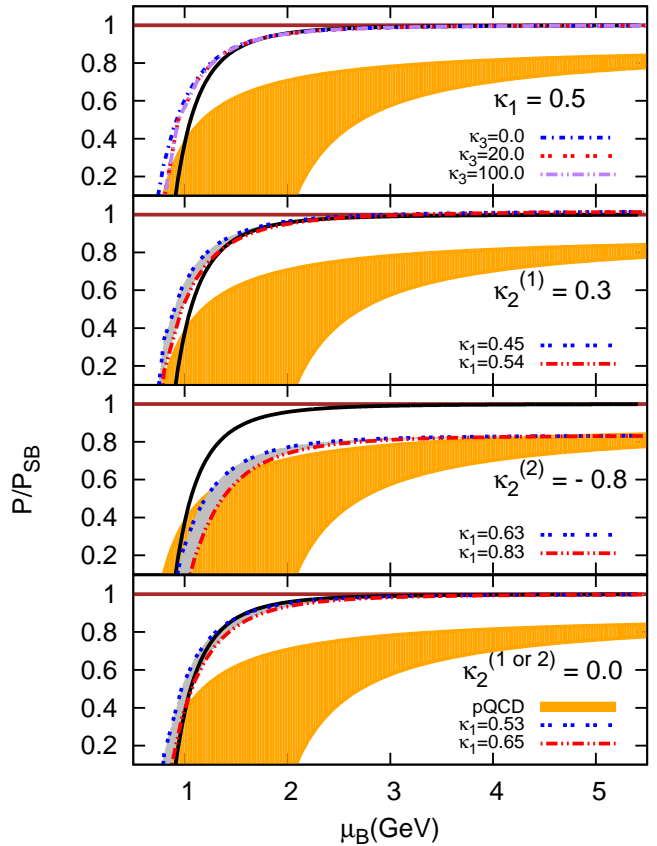, width=9cm}
\caption{\label{fig:pressure} Pressure of SQM  calculated by using several extended CIDDM parameter sets (color dashed line). Solid filled curved line (orange) represents pQCD result\cite{ref:pressure} and the Grey one represents the result from the constraint of the corresponding $\kappa_1$ value. Solid black line represents MIT Bag result with $B^{1/4}=154.4~\rm{MeV}$. The parameter set used in each panel are the same as those used in Fig.~\ref{fig:energy}.} 
\end{figure}
\begin{figure} 
\epsfig{figure=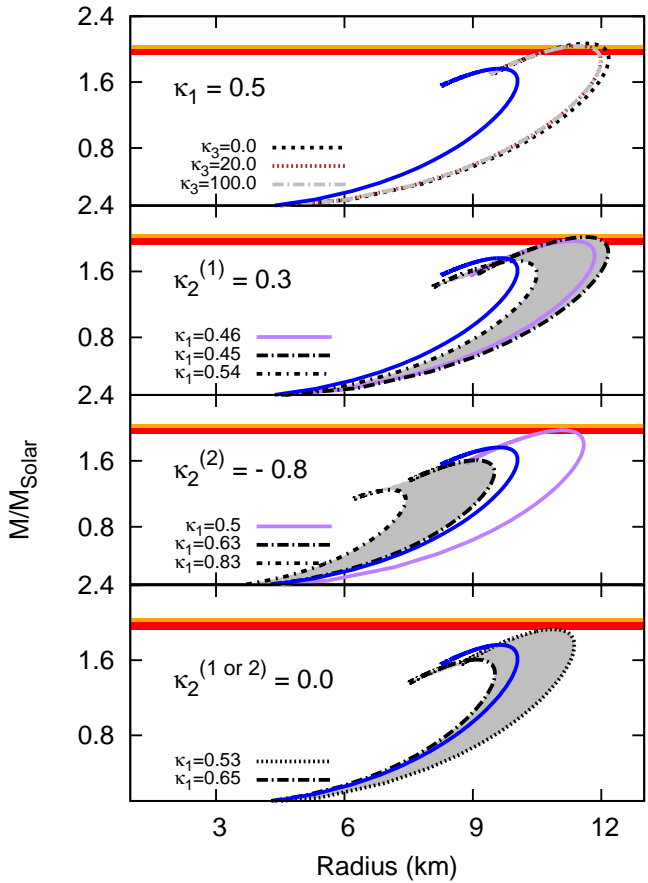, width=9cm}
\caption{Mass-Radius relation for strange star calculated by using several extended CIDDM parameter sets (dashed line). Solid blue line represents MIT Bag result with $B^{1/4}=154.4~\rm{MeV}$. Solid red line represents the mass of pulsar PSR J1614-2230, which is 1.97 $\pm$ 0.04~$M_{\odot}$ and solid orange line for pulsar J0348+0432, which is 2.01 $\pm$ 0.04 $M_{\odot}$. The shaded area (Grey) represents result of the allowed $\kappa_1$ by absolute stability condition constraint. The solid purple is allowed $\kappa_1$ by $2 M_{\odot}$ constraint.  Note: the parameter set $\kappa_1$=0.5 and $\kappa_2^{(2)}$=-0.8 is out of the stability condition constraint. The fixed parameter set used in each panel are the same as those used in Fig.~\ref{fig:energy}.}
\label{fig:radmass}
\end{figure}

From lower panel of Fig.~\ref{fig:pressure}, we can observe the effect of harmonic oscillator parameter. A larger $\kappa_1$ value will lead to lateness with respect to baryon chemical potential for pressure to increase significantly before being saturated at $P$=$P_{SB}$. This happens because a larger pressure at low and moderate densities is needed to suppress the larger interaction generated by the harmonic oscillator term due to increment of $\kappa_1$. From Fig.~\ref{fig:pressure} in second upper and second lower panel, we can see clearly the scalar Coulomb parameter (model I) give different effect to the pressure in high density region compared to vector Coulomb parameter (model II) as estimated by Eq.~(\ref{eq:prediction1}) and Eq.~(\ref{eq:prediction2}). In our calculation, the vector Coulomb parameter gives a more compatible result with the one of pQCD if we put the value of ${\kappa_2}^{(2)}=-0.8$. On the other hand, for the isospin asymmetry term parameter variation, while other parameters are fixed, is shown in upper panel of Fig.~\ref{fig:pressure}. Here $\kappa_1=0.5$, $a=0.8$, $b=0.1 ~\rm{fm}^{3}$ and $\kappa_{2}^{(1~\rm{or}~2)}=0$ are taken. It is obvious that the variation of $\kappa_3$ parameter value from 0 $\le$ $\kappa_3$  $\le$ 100 $\rm {fm}^{3a-1}$  produces only a modest effect in low but negligible effect in high-density regions of EOS.

\begin{figure}
\epsfig{figure=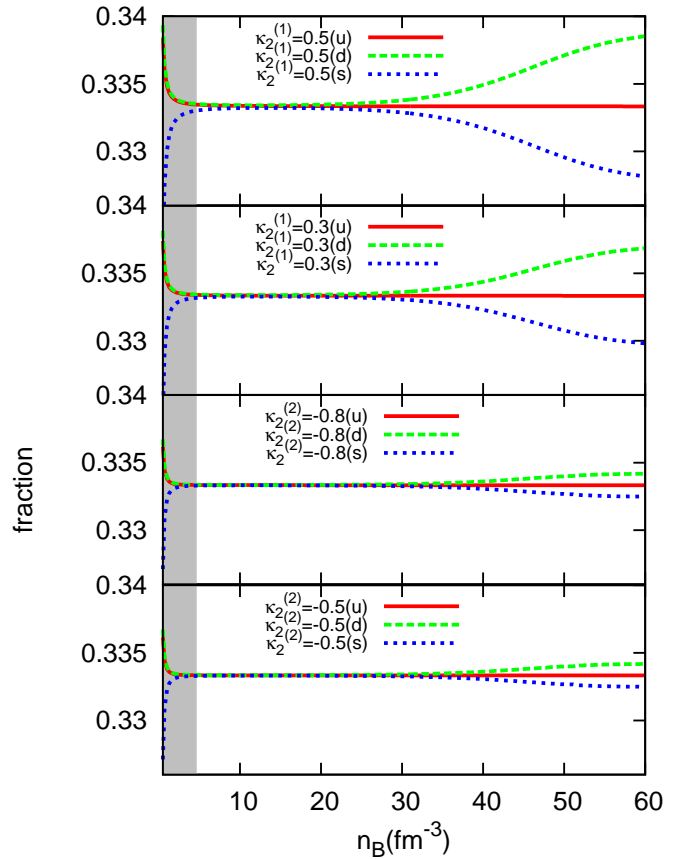, width=9cm}
\caption{Quark's fractions on strange star obtained by using several Coulomb parameter variations. Two lower panel use same fixed parameters which are $\kappa_{1}=0.53$, $\kappa_3=12.7$,$a=0.8$, and $b=0.1~\rm{fm}^{3}$. Another two upper panel use the same as those in lower panel, but here $\kappa_{1}=0.5$ is used instead of $\kappa_{1}=0.53$. The shaded area indicates the range of baryon number density where the fraction give significant effect to mass-radius relation of stars. The range $0.75~\rm{fm}^{-3}<n_B<4.55~\rm{fm}^{-3}$ is deduced mostly by the variation of harmonic oscillator ($\kappa_{1}$) parameter.}
\label{fig:fraksi_coulomb}
\end{figure}

In Fig.~\ref{fig:radmass}, we show the effect of the variation of each parameter of the extended CIDDM model of strange star mass-radius relation. It can be seen clearly from the lower panel of Fig.~\ref{fig:radmass}, the model without additional Coulomb term mostly yields maximum mass less than $2.0~{M}_{\odot}$ except the model with the value of $\kappa_1$ close to the lower bound of the allowed range. While from the second lower panel, the model with the presence of vector Coulomb (model II), where ${\kappa_2}^{(2)}=-0.8$ is used and even the lower bound of $\kappa_1$ is used, yields maximum mass that is still less than $2.0~{M}_{\odot}$. However, it can be seen in the second upper panel, model with the presence of scalar Coulomb (model I), where ${\kappa_2}^{(1)}=0.3$ is used, yields  the maximum masses predicted by some $\kappa_1$ values belonged to the allowed range that match with the $2.0~M_{\odot}$ constraint from pulsar PSR J1614-2230 and PSR J0348+0432. This can be understood because  the  Coulomb parameter of model I tends to stiffen the EOS at high densities while Coulomb parameter of model II tends to soften it. On the other hand, from upper panel, we can see clearly the variation of isospin parameter with the corresponding parameter set yield insignificant effect in mass-radius relation of star. Note that in second lower panel of Fig.~\ref{fig:radmass}, it can be seen in the case  ${\kappa_2}^{(2)}=-0.8$, the  $2.0~M_{\odot}$ can be reached only by using the $\kappa_1$ smaller than those of allowed by absolute stability condition. On the other hand, in the case  ${\kappa_2}^{(1)}=0.3$, the  $M_{\rm max} = 2.0~M_{\odot}$ can be reached by using the $\kappa_1$ that still inside the $\kappa_1$ range which is fulfilling absolute stability condition. 

On this end, it can be concluded that for the models in which their parameters are consistent with absolute stability requirements, the  $M_{\rm max} = 2.0~M_{\odot}$ constraint more prefer the maximum mass prediction of model I (with scalar Coulomb term) than that of model II (with vector Coulomb term). On contrary, the very high densities EOS predicted by  model II (with vector Coulomb) is more compatible with recent pQCD result\cite{ref:pressure}  than that predicted by model I (with scalar Coulomb). It means the presence of more realistic pQCD EOS result i.e., by including the contribution of  charm quarks at very high densities, provides important information to  ensure what kind of Coulomb term of this model is more appropriate. 

In Fig.~\ref{fig:fraksi_coulomb}, the effects of Coulomb terms in the composition of the quarks at high density are shown. The behavior of quark's distribution at high density, on the other hand, in general, depends significantly also on the three parameters of the model used (harmonic oscillator, isospin asymmetry, and Coulomb parameters). In each panel of Fig.~\ref{fig:fraksi_coulomb}, we can see clearly the quark fraction for \textit{up} quark at very high density remains unchanged whatever the type of Coulomb term or $\kappa_1$ value used. However, if we use the vector Coulomb term (model II) in lower and second lower panel, the fraction of \textit{down} and \textit{strange} quarks in this region changes slightly, while for the case of the scalar Coulomb term (model I) in upper and second upper panel, the change in the fraction of \textit{down} and \textit{strange} quark appears more significantly. However, it is clear that the composition of quarks affected by Coulomb term plays less significant role in the region of low up to relative high densities ($0.75 < n_B < 4.55$) where stars probably formed compared to the one at very high densities. 

\section{CONCLUSIONS}
\label{sec_conclu}

In conclusion, for the range of  0 $\le$ $\kappa_3$  $\le$ 100 $\rm {fm}^{3a-1}$ while other parameter are fixed, the parameter variation of isospin asymmetry term produces only a modest effect in binding energy of SQM, mass-radius relation and high density EOS behavior. However, the presence of this term is crucial because the EOS do not fulfill the stability condition for binding energy of u-d quark matter if $\kappa_3$ $\lesssim$ 20. On the other hand, when other parameters are fixed, the parameter variation of harmonic oscillator term, yields significant change not only in mass-radius relation but also the SQM and u-d quark matter binding energies and the prediction EOS of SQM at high densities. It means the role of this term is very crucial for this model. The inclusion of Coulomb terms in CIDDM model could provide more flexibility for the high density EOS prediction of the model in the sense that it can decrease or increase the $\frac{P}{P_{SB}}$ at high densities to be less or more than $1$. The stiffness of EOS of SQM, in general, depends on the parameters range of each term, is controlled by the interplay among the the oscillator harmonic, isospin asymmetry and Coulomb contributions. The predicted EOS by model with Coulomb terms included still can fulfill the absolute  stability condition for SQM.  If the Coulomb term is included, the allowed  $\kappa_1$ by absolute stability condition for SQM can be increased or decreased depending on the kind of Coulomb term used. This is because the scalar Coulomb term tends to stiffen the EOS while vector Coulomb term tends to soften the EOS at high densities.  However, for the models in which their parameters are consistent with SQM absolute stability condition, the  $2.0~M_{\odot}$ constraint more prefer the maximum mass prediction of model with scalar Coulomb term than that of model with vector Coulomb term.  The absolute stability condition of SQM and the mass of 2 $M_\odot$ pulsars, respectively can tightly constrain the parameter of oscillator harmonic, i.e., it yields $\kappa_1 \approx 0.53$ in the case Coulomb term excluded and $0.45 \lesssim \kappa_1 \lesssim 0.46$ in the case scalar Coulomb term with $\kappa_2^{(1)}$=0.3 included. On the contrary, the EOS at high densities predicted by  model with vector Coulomb especially the one with $\kappa_2^{(1)}$ = -0.8 is more compatible with recent pQCD result  than that predicted by model with scalar Coulomb. The fraction of \textit{up},  \textit{down}, and \textit{strange} quarks  are also quite sensitive to parameters of the oscillator, isospin and Coulomb terms used. However, among those terms, the scalar Coulomb term provides the relative significant effect  to stiffen the SQM EOS at very high densities.

\section*{ACKNOWLEDGMENT}
This work has been partly supported by the Research-Cluster-Grant-Program 
of the University of Indonesia, under contract No. 1709/H2.R12/HKP.05.00/2014.
 We acknowledges the support given by Universitas Indonesia.

\end{document}